\documentstyle[preprint,aps,epsf]{revtex}

\newcommand{\CR}{\nonumber \\}
\newcommand{\br}{{\bf r}}

\begin{document}
\draft
\preprint{
\begin{minipage}{30ex}
KUNS-1554,\hfill
APCTP-1999001 
SOGANG-HEP 252/99 \\ {\tt hep-th/9901107}
\end{minipage}
}
\title{Moduli Space Dimensions of Multi-Pronged Strings} 
\author{ 
Dongsu Bak\footnote{dsbak@mach.uos.ac.kr}$^{a}$, 
Koji Hashimoto\footnote{hasshan@gauge.scphys.kyoto-u.ac.jp}$^{b}$, 
Bum-Hoon Lee\footnote{bhl@ccs.sogang.ac.kr}$^{c}$,
Hyunsoo Min\footnote{hsmin@dirac.uos.ac.kr}$^{a}$, 
and
Naoki~Sasakura\footnote{sasakura@gauge.scphys.kyoto-u.ac.jp}$^{b}$}   
\address{a Department of Physics, University of Seoul, Seoul 
130-743 Korea\\ 
b Department of Physics,  Kyoto University , Kyoto 
606-8502 Japan\\ 
c Department of Physics, Sogang University, Seoul 121-742 
Korea}
\tightenlines 
\maketitle 
\begin{abstract} 
{The numbers of bosonic and fermionic zero modes of multi-pronged
  strings are counted in ${\cal N}=4$ super-Yang-Mills theory and
  compared with those of the IIB string theory. We obtain a nice
  agreement for the fermionic zero modes, while our result for the
  bosonic zero modes differs from that obtained in the IIB string
  theory. The possible origin of the discrepancy is discussed. 
} 
\end{abstract}


\section{Introduction}

The recent development of non-perturbative string theories has
provided new powerful tools to understand the supersymmetric gauge
theories. The low energy dynamics of the D-branes are described by the 
supersymmetric gauge theories. The BPS spectrum of the supersymmetric
theory will then correspond to the BPS configurations of strings and
branes ending on the background brane configurations.

Various known properties of the ${\cal N}=4$ $SU(N)$ supersymmetric 
Yang-Mills theory have been studied based on $N$ parallel D3-branes. 
The BPS state spectrum of the massive gauge bosons, monopoles and 
dyons preserving half of the supersymmetry are identified with the
$(p,q)$ strings connecting two separated D3-branes. With more than two 
D3-branes, we can have the string junction configurations\cite{PROSTR} 
that preserve only $1/4$ of the supersymmetries\cite{sen,obergman}. 
The condition for the string junction configurations gives the set of
field equations describing the corresponding BPS states of the gauge
theory\cite{obergman,fraser,hata,lee,kawano,hashimoto}. In addition to
the first order differential equations describing the 1/2 BPS states
of monopoles, the string junction needs a second order equation of the
Gauss law. The field theoretic solutions corresponding to
multi-pronged strings were explicitly constructed for $SU(3)$
theory\cite{hata,lee} and generalized to $SU(N)$
theory\cite{kawano,hashimoto}.

To study the quantum properties of this string junction, one needs 
to understand the zero modes around the classical configurations. As
in the case of monopoles\cite{manton,osborn}, the bosonic zero modes
will correspond to the collective coordinates  of the moduli space,
while the fermionic zero modes correspond to the spin structures of
the supermultiplets. The number of zero modes of monopoles with an
arbitrary gauge group is well known\cite{weinberg}.
 
In this paper, we will count the number of bosonic and fermionic zero
modes of the multi-pronged strings in the $SU(N)$ field theory and
compare it with that of IIB string picture\cite{bergman}.
The number of fermionic zero modes was already discussed for a
specific $SU(3)$ solution in Ref.\ \cite{lee}.
 
In section 2, we briefly describe the BPS equations describing the
string junction and the zero mode equations. In section 3, we count
the zero modes. The equations of the bosonic zero modes consist of
those for the magnetic monopole and one more second order equation. 
Usually, the index, i.e., the number of the zero modes is evaluated by
asymptotic expansion of the field configurations\cite{weinberg}. 
However, for the string junctions, the method of evaluating the index
by the expansion of the electric fields fails. Instead, we will count
the bosonic and fermionic zero modes based on considering the
constraints imposed on the zero modes of multi-monopoles. The detail
of mathematical arguments are in the Appendix A. In section 4, this
counting is shown to be different from that based on the Type II
string theory. In section 5, we summarize our main results, and
indicate future directions.


\section{Multi-Pronged Strings and Their Moduli Space}  

We begin by recapitulating the basic properties of 1/4 BPS states in
the ${\cal N}=4$ $SU(N)$ super-Yang-Mills theory, whose Lagrangian, 
for the bosonic part only, reads
\begin{equation}
\label{e51}
{\cal L} =-{1\over 4 g^2_{\rm YM}}{\rm Tr}\Bigl
[ F^{\mu\nu}F_{\mu\nu}+2 
D_\mu\phi^I D^\mu\phi^I-
2\sum_{I<J}[\phi^I,\phi^J]^2\Bigr] 
\end{equation}
where
$F_{\mu\nu} \equiv \partial_\mu A_\nu-\partial_\nu A_\mu
-i[A_\mu, A_\nu]$,
$D_\mu \phi^I\equiv \partial_\mu\phi^I-i[A_{\mu},\phi^I]$ and the 
indices $I$ and $J$ run from $1$ to $6$. It is well known that, in
this theory, the monopoles, the dyons and the W-particles preserve
half of sixteen supersymmetries of the theory. It is shown  recently
that there may be also 1/4 BPS states that preserve a quarter of the
total supersymmetry. As shown in Refs.\ \cite{obergman,bergman}, these
states describe multi-pronged strings connecting $N$ D3-branes in the
type IIB string picture. Examples of closed-form solutions
corresponding to the pronged strings were found in the ${\cal N}=4$
$SU(N)$ super-Yang-Mills theory\cite{hata,lee,kawano,hashimoto}. We 
shall investigate the number of both bosonic and fermionic zero modes
of the the multi-pronged strings in the ${\cal N}=4$
super-Yang-Mills theory for the  gauge group $SU(N)$. This will be the
first step  to extract the structure or geometry of the moduli space
involved with the multi-pronged strings. Thereby, one is ultimately
interested in obtaining the low-energy effective dynamics of the
multi-pronged strings. 

The Bogomol'nyi bound for the 1/4 BPS states can be found by
considering the energy functional of the ${\cal N}=4$ system, 
\begin{equation}
\label{energy}
M ={1\over 2 g^2_{\rm YM}}\int\! d^3x{\rm Tr}\left[{\bf E}\!\cdot\!
  {\bf E}\! +\! {\bf B}\!\cdot\! {\bf B} +
{\bf D}\phi^I \!\cdot\! {\bf D}\phi^I+D_0\phi^I  D_0\phi^I
-\sum_{I<J}\left[\phi^I,\phi^J\right]^2\right]\,. 
\end{equation} 
Introducing two orthonormal six vectors $e^I$ and $b^I$, we present
the energy equivalently by 
\begin{eqnarray}
\label{energy2}
M&=&\!{1\over 2g^2_{\rm YM}}\!\int\! d^3x {\rm Tr}
\Biggm[
\Bigm|e^I{\bf E}\! +\!
b^I {\bf B} \!-\!
{\bf D}\phi^I \Bigm|^2+\Bigm|D_0\phi^Ie^I\Bigm|^2 
+\Bigm|D_0\phi^Ib^I\!+\!i
\left[\phi^Ie^I, \phi^Jb^J\right]\Bigm|^2
\nonumber\\
&&
\left.
\hspace{110pt} + \Bigm|D_0\phi_{\perp}^I\Bigm|^2 
\!-\!\sum_{I<J}\left[\phi_\perp^I,\phi_\perp^J\right]^2
+2 {\nabla} \!\cdot\!({\bf E}\,\phi^I\! e^I) 
\!+\!2 {\nabla}\! \cdot\!({\bf B}\, \phi^I\!b^I)
 \right]\,, 
\end{eqnarray}
where $\phi_\perp^I$ refers to components of the six-vector
perpendicular to the unit vectors, $e^I$ and $b^I$, and we have used
the Gauss law constraint 
\begin{equation}
\label{gauss}
{\bf D}\!\cdot\! {\bf E}+
i\left[\phi^Ie^I, D_0\phi^Je^J\right] + i
\left[\phi^Ib^I, D_0\phi^Jb^J\right]=0\,, 
\end{equation}
and the Bianchi identity ${\bf D}\!\cdot\! {\bf B}=0$ to perform the
integration by part\footnote{ The Gauss law is, in fact, given by
  ${\bf D}\!\cdot\! {\bf E}+i[\phi^I, D_0\phi^I]=0$ but, with a
  restriction set by the Bogomol'nyi equation $D_0\phi_{\perp}^I=0$,
  one may consistently use the form given here.}.  
This, then, implies that the energy is bounded from below by
\begin{equation}
\label{bound}
g^2_{\rm YM} M \ge  Q_E^I e^I 
+Q_M^I b^I \,, 
\end{equation} 
where we define the charge six vectors as
\begin{eqnarray}
\label{chargevector}
&&Q^I_E \equiv 
\int \!d^3x {\nabla}\!\cdot\!  {\rm Tr}{\bf E}\phi^I   
=Q_E^p h^{Ip},
\nonumber\\
&&Q^I_M \equiv
\int \!d^3 x 
 {\nabla}\! \cdot\!{\rm Tr}{\bf B}\phi^I 
 =Q_M^p h^{Ip}.
\end{eqnarray} 
For the equalities, we used the asymptotic condition,
\begin{equation}
\label{higgs}
\langle\phi^I\rangle = 
h^{Ip} H_p\,,
\end{equation}
where  $H_p$ is the  $N-1$ mutually commuting operators that span the
Cartan subalgebra. The raising and lowering generators $E_{\alpha}$,
\begin{equation}
\label{rasing}
[H_p, E_{\alpha}]=\alpha_p E_\alpha\,,
\end{equation}
are normalized by
\begin{equation}
\label{nomalization}
[E_{\alpha}, E_{-\alpha}]=\alpha_p H_p\,,
\end{equation}
where $\alpha_p$ are the roots. We will choose the simple roots
$\beta_p$ by requiring $h^{Ip} b^I \beta_p >0$ for the maximal
symmetry 
breaking\footnote{We consider only the case of the maximal symmetry
  breaking in the direction of $\phi^Ib^I$ for simplicity. Obviously,
  this condition may be lifted to study the effect of nonabelian
  symmetry breaking.} 
case along $\phi^I b^I$.

The saturation of the bound occurs
if\cite{fraser,hata,lee,kawano,hashimoto}  
\begin{eqnarray}
\label{bogomolnyi}
{\bf D}\phi^I=e^I{\bf E} +
b^I {\bf B}\,,\ \  D_0\phi^Ie^I=0\,, \ \  D_0\phi^Ib^I\!+\!i
[\phi^Ie^I, \phi^Jb^J] =0
\end{eqnarray} 
and 
\begin{eqnarray}
\label{vacuum}
 [\phi_\perp^I,\phi_\perp^J]=0\,,\ \ D_0\phi_\perp^I=0\,.
\end{eqnarray}
There are two types of BPS states that may be classified by 
considering two ${\cal N}=4$ central charges given 
by\cite{fraser}
\begin{eqnarray}
\label{centralcharge}
 Z_\pm=\sqrt{\|Q_E^I\|^2+\|Q_M^I\|^2\pm \|Q_E^I\|\|Q_M^I\|\sin\chi} 
\qquad \Bigl(\chi \in [0,\pi)\Bigr)\,.
\end{eqnarray} 
where  $\chi$ denotes the angle between the charges vectors. 
For $\chi=0$, the state preserves the eight supersymmetries
and these are described field-theoretically by the monopole, 
the dyons and W-particles. With non-vanishing $\chi$, the two
charge-vectors are no longer parallel, and the state preserves only
1/4 of the supersymmetry. We  will be mainly concerned on these 1/4  
BPS states. 

Owing to the tracelessness of the charges, this, in fact, guarantees
the balance of tension of the corresponding multi-pronged
junction. Moreover, one finds a restriction on the electric and the
magnetic charges 
\begin{eqnarray}
\label{torquecon}
  Q_E^{p}h^{Ip} b^I-Q_M^{p}h^{Ip} e^I=0
\end{eqnarray} 
which turns out to be the balance condition of the torque applied
on the associated D3-branes by the pronged string. This restriction
follows from
\begin{eqnarray}
  &&Q_E^{p}h^{Ip} b^I=Q_E^Ib^I =
\int \! d^3 x {\rm Tr} {\bf E}\!\cdot\! 
{\bf D}\phi^I b^I\nonumber\\
&&= \int \!d^3x {\rm Tr} {\bf E}\!\cdot\! 
{\bf B}=\int \!d^3x {\rm Tr} {\bf B}\!\cdot\! 
{\bf D}\phi^I e^I=Q_M^{p}h^{Ip} e^I
\end{eqnarray} 
where we used the Bogomol'nyi equation and the definitions of charges.

The geometric shape of the junction and the meaning of the charges may
be clearly found by fixing our six coordinate system. For later
purpose, we shall denote $\phi^I b^I\equiv A_4$, and $\phi^I e^I\equiv
X$ and further set $\phi_\perp^I=0$ without loss of generality. 

The BPS equations (\ref{bogomolnyi}) are rewritten as 
\begin{eqnarray}
\label{bogo1}
&&{\bf B}={\bf D}A_4\,,\\
\label{bogo2}
&&{\bf E}={\bf D} X \,,\\  
\label{bogo4}
&&D_0 X=0\,,\ \ 
D_0 A_4+{i}[X, A_4]=0\,,
\end{eqnarray}  
with the Gauss law,
\begin{equation}
\label{gauss1}
{\bf D}\!\cdot\! {\bf D}X -[A_4,[A_4,X]]=0\,. 
\end{equation}
Let us now  choose a gauge ${A_0=X}$. Eqs.\ (\ref{bogo2}) and
(\ref{bogo4}) then lead to a relation $\dot{A_m}=\dot X=0$ 
$(m=1,2,3,4)$, which implies any solutions of the BPS equation are
static with this gauge choice. 

Eq.\ (\ref{bogo1}) is the usual BPS monopole equation. Hence the
junction BPS state is a kind of monopole surrounded by W-boson cloud,
which is determined by Eq.\ (\ref{gauss1}). Topological argument leads
to the quantization of the magnetic charge by
\begin{eqnarray}
\label{chargequantization}
Q_M=4\pi\sum_{a=1}^{N-1}m_a \beta^a_p h^{Ip} b^I\,,
\end{eqnarray} 
where the integer $m_a$ counts the number of each fundamental
monopole.

The moduli space of a given multi-pronged string with fixed D-brane
positions is defined by the solution space of the above BPS equation
modulo gauge transformation with fixed vacuum expectation values of
scalars and electric/magnetic charges.

The tangent vectors of the moduli space with the gauge $A_0=X$, will
satisfy the zero-mode equation
\begin{eqnarray}
\label{zero1}
&&
\epsilon^{ijk} {D}_j\delta A_k -{D}_4\delta A^i +
{D}^i \delta A_4\equiv \eta^i_{mn}{D}_m \delta A_n=0\,,\\
\label{zero2}
&&{D}_m {D}_m
\delta A_0  
+2i[{D}_m  {A}_0, \delta A_m]=0\,,\\
\label{gauge2}
&&{D}_m\delta A_m=0\,,
\end{eqnarray}
where we introduced a notation $-i[A_4,(\cdot)]\equiv D_4(\cdot)$ and
the 't Hooft symbol 
\begin{equation}
\label{thooft}
\eta^a
_{mn}=\left\{ 
\begin{array}{ll}
\varepsilon_{amn}\,, & m,n=1,2,3\,,\CR
-\delta_{an}\,, & m=4\,, \CR
\delta_{am}\,, & n=4\,, \CR
0, & m=n=4\,,
\end{array}
\right.
\end{equation}
and Eq.\ (\ref{gauge2}) is the gauge condition.
The zero modes are then normalizable solutions of the above equations
with a norm,
\begin{eqnarray}
\label{norm}
||\delta A ||^2=\int\! d^3x \,
\left(\delta A_0 \delta A_0+
\delta A_m\delta A_m\right)\,.
\end{eqnarray}
The number of the zero modes will agree with the dimensions of the
moduli space or the tangent space at a given point of the moduli
space. In dealing with the above zero-mode equations, we note that a
multi-pronged string of $SU(N)$ can be always embedded into the
$SU(\overline{N})$ theory with a larger $\overline{N}$. Then in the
$SU(\overline{N})$ theory, the field components of the multi-pronged
solution other than the $SU(N)$ component are zero by
construction. Then with this background solution, the  fluctuations of
components other than the $SU(N)$ component will be dynamically
decoupled in the above zero-mode equations and, hence, can be
trivially set to vanish from the beginning. Hereafter, we shall choose
such an $SU(N)$ subgroup with  the minimum rank, where one can embed
the pronged string, and work within such $SU(N)$. This $N$ is then the
number of D-branes where the prongs end, which will be
denoted by $\widetilde{N}$.

To expose the structure of the low-energy effective Lagrangian
resulted from the moduli-space approximation, let us suppose the zero
modes are given by $\delta_s A \,\,\, (s=1,2,\cdots, {\rm
  \#zero\!\!-\!\!mode})$. Let us further denote the moduli-space
element as $A(\br;\xi)$ with $\xi^s$ being the coordinate of the
moduli space.

Inserting the solutions to the Lagrangian with time-dependent 
moduli-coordinates, one finds that
\begin{eqnarray}
\label{efflag2}
L_{\rm eff}={1\over 2 g^2_{\rm YM}}\int\! d^3x \, {\rm Tr}\Bigl[
\dot{A}_0 \dot{A}_0+\dot{A}_m\dot{A}_m-
2\partial_i( A_0 \dot{A}^i)\Bigr]\,,
\end{eqnarray}
where we have used the gauge condition $A_0=X$ and have dropped a
constant term. The effective Lagrangian may be rewritten, to the
quadratic order in velocities, as
\begin{eqnarray}
\label{efflag3}
L_{\rm eff}={1\over 2} g_{ss'}(\xi)\dot{\xi}^s\dot{\xi}^{s'}
-{\cal A}_s(\xi)\dot{\xi}^s\,,
\end{eqnarray}
with an appropriate gauge transformation,
\begin{eqnarray}
\label{gaugetr}
 \dot A\rightarrow \dot\xi^s(\partial_s A-D\epsilon_s)\equiv 
\dot\xi^s \delta_s A\,,
\end{eqnarray}
that insures the gauge condition (\ref{gauge2}) of the zero mode.
The metric and the vector potential can be expressed, in terms of the
zero mode:
\begin{eqnarray}
\label{metric}
&&g_{ss'}(\xi)=
{1\over g^2_{\rm YM}}\int \! d^3x \, {\rm Tr}
\left[\delta_s {A}_0 \delta_{s'}{A}_0+
\delta_s{A}_m\delta_{s'}{A}_m\right]\,, \\
\label{potential}
&&{\cal A}_s(\xi)={1\over g^2_{\rm YM}}\int_{r=\infty}\!\!\!dS^i \,
{\rm Tr} A_0 \delta_s A_i=
\int\! d\Omega \,h^I_p e^I \lim_{r\rightarrow \infty}
\left[r^2 {\rm Tr}\left(\hat{r}^i \delta_s A^i  H_p
\right)\right] \,,
\end{eqnarray}
where, in the last equation, we have used the normalizability
condition $\delta_s A_i=O(1/r^2)$ for a large  $r$. The vector
potential term in Eq.\ (\ref{efflag3}) is in fact a total time
derivative term. This can be easily seen from the last term of Eq.\ 
(\ref{efflag2}), where $A_0$ can be replaced by  its asymptotic value.
Although this total time derivative term does not affect the classical
dynamics of the moduli space, it is relevant in its quantum version
especially when some of the directions of the moduli space are
compact. The motion in these compact directions is indeed involved
with the quantization of the electric charges.


\section{Counting the Number of Zero Modes}

The equations of the bosonic zero modes and the gauge fixing condition
are given by Eqs.\ (\ref{zero1}), (\ref{zero2}) and (\ref{gauge2}). In
this section we shall analyze the number of the normalizable solutions
of these equations. Here and below, all vector potential and the
covariant derivative denote, respectively, the background  solutions
of pronged strings and the covariant derivative  with respect to the
background. The normalizable solutions of Eqs.\ (\ref{zero1}) and
(\ref{gauge2}) are the zero modes of a BPS monopole. The problem here
is whether the equation (\ref{zero2}) gives a normalizable solution
for $\delta A_0$ or not. As discussed in appendix
(\ref{app:normalizable}), the condition that the solution of Eq.\
(\ref{zero2}) be normalizable is given by
\begin{equation}
\int\! d^3x\, {\rm Tr}
\left(\Lambda_a\left[ D_m A_0, \delta A_m\right]\right) =0\,,
\label{eq:connor}
\end{equation}
where the trace is over the color indices and $\Lambda_a\
(a=1,\cdots,\widetilde{N}\!-\!1)$ are the zero modes of the operator 
$-{D_m}D_m$. From the junction BPS equation, $D_mD_m A_0=D_m D_m
A_4=0$ holds. They are independent for a 1/4 BPS state, and hence they 
are two of $\Lambda_a$. As shown in appendix (\ref{app:normalizable}),
the conditions (\ref{eq:connor}) are equivalent to the condition that
the electric charges should not change under the infinitesimal changes
of the configuration.

Apparently these conditions seem to give $\widetilde{N}\!-1\!$
conditions on the tangent moduli space of the monopole, but this is
not. In fact, one of the conditions 
\begin{equation}
\int \!d^3x\, {\rm Tr}(A_4 [D_m A_0, \delta A_m]) =0
\label{eq:idcon}
\end{equation} 
is satisfied for any monopole zero mode $\delta A_m$. To show this, we
can use an identity 
resulting from the simple fact that the magnetic charges do not change
under the infinitesimal change. As shown in appendix
(\ref{app:normalizable}), this implies that 
\begin{equation}
\int\! d^3x\, {\rm Tr} (A_0 [D_m A_4, \delta A_m]) =0\,.
\end{equation} 
One can easily show that the partial integration of this equation
gives Eq.\ (\ref{eq:idcon}). (Alternatively, one may understand Eq.\
(\ref{eq:idcon}) in terms of the torque balance identity
(\ref{torquecon}) by $ \delta {\rm Tr}[A_4(r\!=\!\infty)\,Q_E ] = 
 \delta {\rm Tr}[A_0(r\!=\!\infty)\,Q_M ]$ and $\delta Q_M =0$.)
Hence the number of the constraints is in fact $\widetilde{N}\!-\!2$.
The number of the bosonic zero modes of the monopole BPS equations is
given by $\#{\rm  monopole} \times 4$ \cite{weinberg}, where 
$\#{\rm  monopole}$ is the total number of the fundamental monopoles, 
$\sum_a m_a$ (see Eq.\ (\ref{chargequantization})). Therefore, taking
into account the constraints, the total number of the bosonic zero
modes (BZM) of the junction solution is given by 
\begin{equation}
\# {\rm BZM}= \#{\rm  monopole}\,\times 4-\widetilde{N}+2\,.
\label{eq:totbzm}
\end{equation}

Let us now discuss the number of the fermionic zero modes of the 
junction solution. The equations for the fermionic zero modes are
given by
\begin{eqnarray}
\label{eq:ferzeromin}
{\cal D}^{(-)}\psi^{(-)}&=&\left( \begin{array}{ccc}
-2D_0 & \tau_m^- D_m \\
\tau_m^+ D_m & 0 
\end{array}
\right) \psi^{(-)} =0\,,\\
{\cal D}^{(+)}\psi^{(+)}&=&\left( 
\begin{array}{ccc}
0 & \tau_m^- D_m \\
\tau_m^+ D_m & -2D_0 
\end{array}
\right) \psi^{(+)} =0\,, 
\label{eq:ferzeroplus}
\end{eqnarray}
where $D_0$ is understood as $D_0 (\cdot) =-i[A_0, (\cdot)]$ 
and the $2\times 2$ matrices is defined by  $\tau_m^-=(\sigma^i, i)$
and $\tau_m^+=(\sigma^i, -i)$ with the Pauli matrices $\sigma^i$. To
first analyze the equation (\ref{eq:ferzeroplus}), we decompose it by 
\begin{equation}
\psi^{(+)}=\left(\begin{array}{c} \psi^{(+)}_1 \\ \psi^{(+)}_2
  \end{array}\right).
\end{equation}
Then the equation (\ref{eq:ferzeroplus}) is 
\begin{eqnarray}
\tau_m^- D_m \psi^{(+)}_2 &=&0\,, \\
\tau_m^+ D_m \psi^{(+)}_1- 2 D_0 \psi^{(+)}_2 &=&0\,. 
\end{eqnarray}
Since the operator $\tau_m^- D_m$ has no zero modes, the solution is 
given by that $\psi^{(+)}_2=0$. Hence $\tau_m^+ D_m \psi^{(+)}_1=0$,
and $\psi^{(+)}_1$ is just the fermionic zero modes of a BPS monopole.
Thus the number\footnote{We count the number in real component. Hence 
  $\psi$ and $i\psi$ are counted as distinct solutions.} of the
fermionic zero modes resulting from Eq.\ (\ref{eq:ferzeroplus}) is
given by $\# {\rm monopole}\, \times 4$. 

With a similar decomposition to two-component spinors, Eq.\
(\ref{eq:ferzeromin}) can be rewritten as
\begin{eqnarray}
\tau_m^+ D_m \psi^{(-)}_1 &=&0\,,
\label{eq:ferzerominone} \\
\tau_m^- D_m \psi^{(-)}_2- 2 D_0 \psi^{(-)}_1 &=&0\,.
\label{eq:ferzeromintwo}
\end{eqnarray}
The equation (\ref{eq:ferzerominone}) just gives the fermionic zero
modes of a BPS monopole and the number of the solutions is $4\times\#
{\rm monopole}$. Let us discuss the normalizability of $\psi^{(-)}_2$
resulting from Eq.\ (\ref{eq:ferzeromintwo}). First act an operator
$\tau_m^+ D_m$ 
on Eq.\ (\ref{eq:ferzeromintwo})\footnote{The following discussions
  are obscured   by the existence of the zero modes of the operator
  $\tau_m^+ D_m$. However, using the explicit known relations between
  the fermionic and bosonic zero modes of a monopole, one can show the 
  same result given below. N.S. would like to thank S.\ Imai for
  clarifying this point.}. 
Using the BPS equation, we obtain 
\begin{equation}
D_m D_m \psi^{(-)}_2 -2 \tau_m^+ D_m D_0 \psi^{(-)}_1=0\,.
\end{equation} 
This equation looks very similar to the bosonic one (\ref{zero2}).
To obtain a normalizable solution of $\psi^{(-)}_2$, the following
constraints on $\psi^{(-)}_1$ must be satisfied:
\begin{equation}
\int\! d^3x\, {\rm Tr}\left(  \Lambda_a \tau_m^+ D_m D_0
  \psi^{(-)}_1\right)=0\,, 
\label{eq:idnorferzero}
\end{equation}
where the trace is over the color indices. As similar to the bosonic
case, one of the constraints,
\begin{equation}
\label{eq:constfermion}
\int\! d^3x\, {\rm Tr}\left( A_4 \tau_m^+ D_m D_0
  \psi^{(-)}_1\right)=0\,, 
\end{equation}
is identically satisfied by the monopole fermionic zero modes. 

To show this, let us start with Eq.\ (\ref{eq:ferzerominone}). 
Applying the
operator $\tau_m^- D_ m$ on it, one obtains 
$(D_m D_m+i\bar \eta^a_{mn}\sigma_a D_m D_n) \psi^{(-)}_1=0$ (The
symbols $\bar \eta^a_{mn}$ differ from $\eta$ by a change in the sign
of $\delta$ in the definition (\ref{thooft})).
Using further the BPS equations, we obtain
\begin{equation}
(D_mD_m - 2 i \tau_m^+ D_m D_4 )\psi^{(-)}_1=0\,.
\end{equation}
Hence the normalizability of the fermionic zero mode $\psi^{(-)}_1$
reads
\begin{figure}[htdp]
\begin{center}
\epsfxsize=90mm
\epsfbox{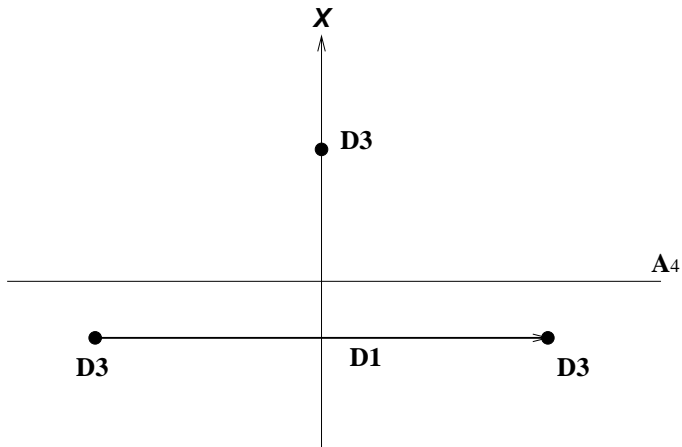}
\caption{The configuration of D-string ending on D3-branes is
  depicted. The figure is for magnetic charges $(m_1,m_2)=(1,1)$ in
  $SU(3)$ case. The moduli-space is four dimensional.}
\label{dstring}
\end{center}
\end{figure}
\begin{equation}
\int\! d^3x\, {\rm Tr}\left(
\Lambda_a \tau_m^+ D_m D_4 \psi^{(-)}_1\right)=0\,.
\label{eq:norferzero}
\end{equation}
By the partial integration of Eq.\ (\ref{eq:norferzero}) with the 
substitution $\Lambda_a=A_0$, we obtain Eq.\ (\ref{eq:constfermion}).

Thus the constraints (\ref{eq:idnorferzero}) give $4(\widetilde{N}-2)$
constraints on the fermionic BPS zero modes. Thus the total number of
the fermionic zero modes (FZM) is given by
\begin{equation}
\# {\rm FZM}= \#{\rm  monopole}\, \times 8 -4(\widetilde{N}-2)\,.
\label{eq:totalferzero}
\end{equation}
In the next section, we will compare the numbers of the zero modes,
Eqs.\ (\ref{eq:totbzm}) and (\ref{eq:totalferzero}), with those
derived from the type IIB description.

Finally we mention the reason why we should take $\widetilde{N}$
instead of $N$ in our formulae for the numbers of the zero modes.
Let us consider a D-string of $SU(3)$ theory in
FIG. \ref{dstring}. The zero mode equations take the same form as
Eqs.\ (\ref{zero1}), (\ref{zero2}) and (\ref{gauge2}) with $D_m
A_0=0$. Hence the solution of Eq.\ (\ref{zero2}) is simply $\delta
A_0=0$. The others possess $2\times 4$ zero-modes because the D-string
is regarded as two fundamental monopoles in Eqs.\ (\ref{zero1}) and
(\ref{gauge2}). This is incorrect because we know from the beginning,
there are only four moduli  degrees of freedom around this
configuration. This failure of the zero modes analysis may be
understood as follows. Around the D-string configuration, the
field-theoretic potential is too flat to capture the correct number of 
zero modes by just considering the linearized fluctuations of
Eqs.\ (\ref{zero1}), (\ref{zero2}) and (\ref{gauge2}). The higher
order analysis will show that the the relative motions of the two
fundamental monopoles drops out of the moduli space, which leaves just
four overall translation degrees. 

However, our formula, Eqs.\ (\ref{eq:totbzm}) and
(\ref{eq:totalferzero}), for the zero modes  are still valid for the
D-string because we are using the minimal embedding of solutions and,
hence, $\widetilde{N}=2$. 

\section{Comparison with IIB String Theory}

Our result of the number of the bosonic zero modes does not agree with
that from the IIB string theory. To illustrate, let us consider the
simplest case of a tree three-pronged string with two-form charges
(1,0), (0,1) and $(-1,-1)$. The magnetic charge of the corresponding
field theory solution in $SU(3)$ is $(m_1,m_2)=(1,1)$. Thus the
magnetic part has two fundamental 
monopoles, and the result (\ref{eq:totbzm}) tells
that there are seven bosonic zero modes. This number looks quite odd,
and moreover this does not agree with the IIB result that is just
three for the present case\cite{bergman}. However this is a very
natural result from the explicit junction solution discussed in Ref.\
\cite{lee}. Their solution is composed of two monopole cores 
which are surrounded by clouds of W-bosons, i.e. electric charges. The
monopole part has eight bosonic zero modes which are composed of the
zero modes associated to three translations, two gauge, two relative
orientations, and one relative distance.  The electric part is
determined by the monopole part when the vacuum expectation values of
the scalar fields take given fixed values. Especially, the electric
charges of the junction solution are determined by the relative
distance, while the magnetic charges are just topological numbers and
stable. Since the electric charges appear in the asymptotic $1/r$
behavior of the junction solution, the change of the relative distance
is not normalizable. Thus we should keep the relative distance
fixed. The other monopole zero modes  are naturally expected to be
normalizable, and hence there are seven bosonic zero modes in the
junction solution.

\begin{figure}
[htdp]
\begin{center}
\epsfxsize=120mm
\epsfbox{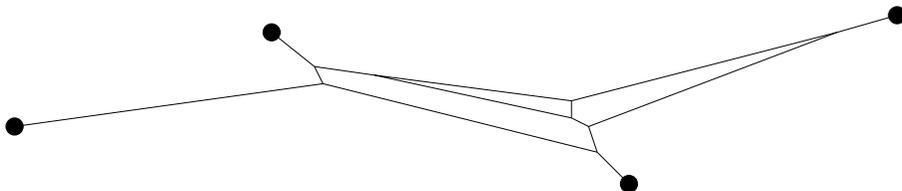}
\caption{The configuration of pronged strings ending on D3-branes
  which corresponds to our field theory analysis.  The long strings
  have R-R charge one and small NS-NS charges. The figure is for
  magnetic charges $(m_1,m_2,m_3)=(1,3,2)$ in $SU(4)$ case.}
\label{fig:stringcon}
\end{center}
\end{figure}

On the other hand, we find a nice agreement for the fermionic part.
Let us consider a monopole configuration with magnetic charges
$(m_1,m_2,\cdots,m_{\widetilde N-1})$. The corresponding string
configuration is such that $\widetilde N$ D3-branes are aligned on a
line and that the $a$-th and the $(a+1)$-th D3-brane are connected by
$m_a$ D-strings $(a=1,\cdots,\widetilde{N}-1)$. This configuration may
be regarded as the string configuration corresponding to a junction
solution in the limit of vanishing electric charges. To recover from
the limit, let us now add small NS-NS charges on each string. We do
not take care of the quantization of the NS-NS charges, since our
treatment in the field theory is just classical and does not care
about the quantization of the electric charges. Then the configuration 
of the string is deformed, for example, to the one in  FIG.\
\ref{fig:stringcon}, where there are $m_a-1$ internal loops between
the $a$-th and $(a+1)$-th D3-branes. Although the diagram changes if
the assignment of the small NS-NS charges is changed, the number of
loops do not change. The total number of the loops in the diagram is
given by
\begin{equation}
\sum_{a=1}^{\widetilde{N}-1} (m_a-1)=\# {\rm monopole} -
(\widetilde{N}-1)\,. 
\end{equation}

The zero mode analysis in the IIB framework was done in Ref.\
\cite{bergman}. 
The result is 
\begin{equation}
\# {\rm FZM(IIB)}=8 F_{\rm int} + 4 E_{\rm ext}\,,
\end{equation}
where $F_{\rm int}$ denotes the number of internal loops (faces) of
the string diagram and $E_{\rm ext}$ is the number of the external
strings. Thus, applying to the present case, we obtain
\begin{equation}
\# {\rm FZM(IIB)}=8\left(\# {\rm monopole}-(\widetilde{N}-1)\right) +
4 \widetilde{N} =
\# {\rm monopole}\,\times 8 -4\widetilde{N} +8\,,
\end{equation}
which agrees with Eq.\ (\ref{eq:totalferzero}).

Finally, the discrepancy  of the bosonic zero modes from the IIB
string picture might be understood as follows. The  expression
(\ref{eq:totbzm}) of the bosonic zero modes  can be written as 
\begin{equation}
\label{disc}
\# {\rm BZM}=\Bigl(2(\# {\rm monopole}-1)+\# {\rm monopole}\Bigr) +
  \# {\rm BZM(IIB)}\,, 
\end{equation}
where the last term is the IIB result\cite{bergman},
\begin{equation}
\# {\rm BZM(IIB)}=F_{\rm int} +3\,.
\end{equation}
It is intriguing to note that the discrepancy (\ref{disc}) agrees with
the number of the compact directions of the monopole moduli space,
i.e. there are $2(\# {\rm monopole}-1)$ relative spatial orientations
among the fundamental monopoles and one $U(1)$ gauge direction per
each fundamental monopole. It might be expected that, when the
junction BPS state is treated quantum mechanically, the wave function 
prevails the compact directions and these directions do not appear as
the moduli of the state.


\section{Conclusion}

In this paper, we nonperturbatively identified the numbers of the
bosonic and fermionic zero modes of the multi-pronged strings in the
context of the ${\cal N}=4$ super-Yang-Mills theory. The bosonic zero
modes differ from the IIB string picture, but the fermionic zero modes
are matching with those in the IIB string picture.

The discrepancy is due to the softness of the field-theoretic
configurations. Namely the monopoles of the multi-pronged strings in
the field theory can take a relative motion in the parallel space of
the D3-branes, whereas the corresponding degrees in the IIB picture
cannot be permitted. In the case of the minimal three-pronged strings, 
the number of bosonic zero modes are seven while there are twelve
fermionic zero modes. On the ground of the remaining supersymmetries
of the system, the natural number of bosonic degrees would be even due
to the complex structure of the remaining supersymmetry. We expect
that the analysis of detailed moduli dynamics may be helpful in
resolving this issue. The comparison with the M-theory
result\cite{sugimoto} or the D-string worldsheet approach\cite{hashi}
would also be interesting. 

The dynamics of the moduli space is in itself of importance,
especially in related with the quantizations of the electric charges.
The supersymmetric quantum mechanics of the moduli space has been
constructed in  case of monopoles\cite{gauntlett,blum}. Our work can
be used in the identification of the supersymmetric quantum mechanics
for the multi-pronged strings. As is also done for the monopoles and
dyons\cite{clee}, the response analysis of the multi-pronged strings
to the excitations of unbroken gauge fields will clarify most of
leading physical processes around the multi-pronged strings. These
require further studies.
 
\vspace{60pt}
\begin{acknowledgements} 
K.\ H.\ and N.\ S.\ thank APCTP for hospitality, where a 
part of this work was done. They would like to thank K.\ Lee, S.\
-J.\ Rey and P.\ Yi for interesting discussions at the stay. D. B.
was supported in part by BSRI Program under BSRI 98-015-D00061, KOSEF
Interdisciplinary Research Grant 98-07-02-07-01-5, KOSEF through
SNU-CTP, and UOS Academic Research Program.
B.-H.\ L. was supported in part by BSRI Program under BSRI 98-2414,
and KOSEP  through SNU-CTP. 
N.\ S.\  was supported in part by Grant-in-Aid for Scientific Research
from Ministry of Education, Science and Culture (\#09640346) and
Priority Area: ``Supersymmetry and Unified Theory of Elementary
Particles'' (\#707). K.\ H.\ was supported in part by Grant-in-Aid for
JSPS fellows.
\end{acknowledgements} 

\clearpage
\appendix

\section{The Normalizability Condition}
\label{app:normalizable}

In this appendix, we will show that the condition of getting a
normalizable $\delta A_0$ from Eq.\ (\ref{zero2}) is given by
Eq.\ (\ref{eq:connor}). 

To simplify the expressions, we take a gauge where the vacuum
expectation value of $A_4$ is expressed in a diagonal form\footnote{We 
  cannot take this gauge globally. In the following discussions, we
  just need to take this gauge for a certain solid angle less than
  $4\pi$ outside a sphere of sufficiently large radius, since the
  solid angle can be chosen arbitrary.}.
We assume also that the diagonal entries take general distinct values.
Then, since the massless fields are associated only to the diagonal 
entries, the asymptotic behavior of the solutions to the equation 
$D_mD_m \Lambda_a=0$  should take the form
\begin{equation}
\Lambda_a= h_a^{(0)}+{h_a^{(1)}\over r}+
O\!\left({1\over r^2}\right)\,,
\label{eq:infexplam1}
\end{equation}
where $h_a^{(0,1)}$ are diagonal matrices. We assume that there are 
$\widetilde N-1$ solutions to this equation $(a=1,\cdots,
\widetilde{N}-1)$,
and that the $h_a^{(0)}$ span the Cartan subalgebra\footnote{This
is explicitly shown for the solutions in
\cite{hata,lee,kawano,hashimoto}.}. 
Note that $\Lambda_a$ includes $A_0$ and $A_4$, 
and we denote $\Lambda_1=A_0$ and
$\Lambda_2=A_4$. The diagonal entries of the $h_1^{(1)}$ are the
electric charges of the junction solution, while those of $h_2^{(1)}$
are the magnetic charges. Since the vacuum expectation values are
fixed, the infinitesimal changes caused by the changes of the moduli
parameters of a monopole should be in the form
\begin{equation}
\delta A_0 = {\delta h_1^{(1)}\over r} + O\!\left({1\over
    r^2}\right)\,,  
\qquad
\delta A_4 = {\delta h_2^{(1)}\over r} + O\!\left({1\over
    r^2}\right)\,. 
\label{eq:infexplam}
\end{equation}
Since the magnetic charges are topological and do not change, $\delta
A_m=O(1/r^2)$. We define generalized electric central charges by 
\begin{eqnarray}
Q^E_{\Lambda_a}\equiv\int_{r=\infty} \!\!\!dS_i\, {\rm Tr}(\Lambda_a
E_i)=\int_{r=\infty} \!\!\!dS_i\, {\rm Tr}(\Lambda_a
D_iA_0)= {\rm Tr}\left(h_a^{(0)}h_1^{(1)}\right)\,.
\end{eqnarray}
Under the assumption that the $h_a^{(0)}$ span the Cartan subalgebra,
the invariance of the electric charges is equivalent to the invariance 
of these central charges.
Since $h_a^{(0)}$ is fixed in the zero-mode analysis, the deformation
of these central charges by the presence of the zero modes is  
\begin{eqnarray}
\delta Q^E_{\Lambda_a}=\int_{r=\infty} \!\!\!dS_i\, {\rm Tr}[\Lambda_a 
\delta (D_i A_0)]=\int \!d^3x\, {\rm Tr} 
\left[D_m\Lambda_a \delta(D_m A_0)
+\Lambda_a D_m\delta(D_m A_0)\right]\,.   
\label{deform}
\end{eqnarray}
Using the above asymptotic behaviors together with a little further
manipulation of Eq.\ (\ref{deform}), one obtains 
\begin{eqnarray}
\delta Q^E_{\Lambda_a}=\int\! d^3x\, {\rm Tr} [\Lambda_a D_m D_m
\delta A_0] 
=-2i\int\! d^3x\, {\rm Tr}\Bigl(\Lambda_a [D_m A_0,\delta
A_m]\Bigr)\,,
\label{edeform}
\end{eqnarray}  
where we have used Eq.\ (\ref{zero2}) for the last
equality. Similarly, defining 
\begin{eqnarray}
Q^M_{\Lambda_a}\equiv\int_{r=\infty} \!\!\!dS_i\, {\rm Tr}(\Lambda_a
B_i)=\int_{r=\infty} \!\!\!dS_i\, {\rm Tr}(\Lambda_a
D_iA_4)= {\rm Tr}\left(h_a^{(0)}h_2^{(1)}\right)\,,
\end{eqnarray}
the deformation by the zero mode can be expressed as
\begin{eqnarray}
\delta Q^M_{\Lambda_a}
=-2i\int\! d^3x\, {\rm Tr}\Bigl(\Lambda_a [D_m A_4,\delta A_m]
\Bigr)\,,   
\label{mdeform}
\end{eqnarray} 
which is in fact automatically vanishing due to $\delta Q_M=0$ or
$\delta h_2^{(1)}=0$. Thus, recalling the fact that there are
$\widetilde{N}-1$ degrees of freedom of the electric charges and from
Eq.\ (\ref{zero2}), the equation (\ref{eq:connor}) is equivalent to
the condition that 
the electric charges do not change under the infinitesimally small
changes of the monopole moduli. Since the electric charges appear in
the asymptotic $1/r$ behavior of $\delta A_0$, this is a necessary
condition for the infinitesimal change to be normalizable under the
measure $\int \!d^3x\, {\rm Tr}({\delta A_0}^2)$. This necessary
condition becomes a sufficient condition if the order next to $1/r$ is
$1/r^2$ as in the expansion (\ref{eq:infexplam1}) and
(\ref{eq:infexplam}). In the spherically symmetric solution discussed
in Refs.\ \cite{hata,kawano,hashimoto}, the next order is
exponentially damping. In general non-spherical cases\cite{lee}, the
next order is expected to behave as $1/r^2$ from dipole contributions.

There is another way to see the condition (\ref{eq:connor}). Since the
operator $-D_m D_m$ is a semi positive definite Hermitian operator,
one may expand $\delta A_0$ in terms of the eigenfunctions. The
equation (\ref{zero2}) is now 
\begin{equation}
PC(P;\Omega)-2i \int\! d^3x\, {\rm Tr}(f(P;\Omega)^\dagger 
[D_m A_0, \delta A_m]) =0 ,
\label{eq:expinfgauss}
\end{equation}
where $f(P;\Omega)$ denotes the eigenfunction with eigenvalue $P$ with
$\Omega$ parameterizing the degeneracies of the eigenfunctions with
the same eigenvalue, and $\delta A_0$ is expanded as $\sum_{P,\Omega}
C(P;\Omega)f(P;\Omega)$. If the second term of Eq.\
(\ref{eq:expinfgauss}) is 
non-zero at $P=0$, the $C(P;\Omega)$ will behave in $1/P$ near $P=0$
(We assume that the eigenvalues of the operator $-D_m D_m$ exist
continuously around $P=0$). This behavior may violate the
normalizability. Therefore the condition that the second term vanishes 
for $P\rightarrow 0$ is related to the normalizability. This is the
condition (\ref{eq:connor}). But to conclude this we need the measure
near $P=0$ 
from more knowledge on the spectrum of the eigenvalues and the
eigenvectors.

\newcommand{\J}[4]{{\sl #1} {\bf #2} (#3) #4}
\newcommand{\andJ}[3]{{\bf #1} (#2) #3}
\newcommand{\AP}{Ann.\ Phys.\ (N.Y.)}
\newcommand{\MPL}{Mod.\ Phys.\ Lett.}
\newcommand{\NP}{Nucl.\ Phys.}
\newcommand{\PL}{Phys.\ Lett.}
\newcommand{\PR}{Phys.\ Rev.}
\newcommand{\PRL}{Phys.\ Rev.\ Lett.}
\newcommand{\PTP}{Prog.\ Theor.\ Phys.}
\newcommand{\hep}[1]{{\tt hep-th/{#1}}}


\end{document}